\documentstyle[aps,twocolumn,epsf,prb,floats]{revtex}
\def\geqap{\,\raise 2pt \hbox{$>\kern-11pt \lower 5pt \hbox{$\sim$}$}\,}
\def\leqap{\,\raise 2pt \hbox{$<\kern-10pt \lower 5pt \hbox{$\sim$}$}\,}
\begin{document}
\draft
\twocolumn[\hsize\textwidth\columnwidth\hsize\csname @twocolumnfalse\endcsname
\title{Resonant Inelastic X-ray Scattering from Charge and Orbital Excitations in Manganites}
\author{H.~Kondo, S.~Ishihara, and S.~Maekawa}
\address{Institute for Materials Research, 
Tohoku University, Sendai 980-8577, Japan}
\date{\today}
\maketitle
\begin{abstract}
We present a theory of the resonant inelastic x-ray scattering (RIXS) 
to study electronic excitations in orbital ordered manganites. 
The charge and orbital excitations of the Mn $3d$ electron are caused 
by the Coulomb interactions in the intermediate scattering state. 
The scattering cross section is formulated by the Liouville operator method  
where the local and itinerant natures of the excitations are taken into account on an equal footing. 
As a result, the cross section is expressed by the charge and orbital correlation functions 
associated with local corrections. 
The RIXS spectra are calculated numerically as functions of momentum and polarization of x ray. 
Through the calculations, 
we propose that RIXS provides a great opportunity to study 
the unique electronic excitations in correlated electron systems with orbital degeneracy.  
\end{abstract}
]
\narrowtext
%
%
\section{introduction}
Since the discovery of High-Tc superconducting cuprates, 
electronic structure of transition-metal oxides has been reinvestigated 
from a view point of electron correlation. \cite{imada} 
An important consequence of the strong electron correlation 
is a variety of electronic phases and 
unique elementary excitations attributed to the internal degrees of freedom of an electron.  
The most exotic example is the spin-charge separation in one-dimensional metals 
where the collective spin and charge excitations behave independently. 
Even in the two- and three-dimensional systems, 
it is recognized that characteristic momentum and energy dependence of 
electronic excitations plays crucial roles in anomalous metallic states 
near the metal-insulator transition. 
\par
In addition to the spin and charge degrees of freedom, 
manganites with perovskite structure $A_{1-x}B_x$MnO$_3$ ($A$=La, Pr, Nd, Sm, 
$B$=Sr, Ca, Ba) have 
orbital degree of freedom; \cite{tokura}
one of the doubly degenerate $e_g$ orbitals, i.e. 
$3d_{3z^2-r^2}$ and $3d_{x^2-y^2}$ orbitals, is occupied by an electron in a Mn$^{3+}$ ion.
LaMnO$_3$ is an antiferromagnetic insulator, \cite{wollan} as is La$_2$CuO$_4$. 
Moreover, the orbital degree of freedom shows the long range ordering 
accompanied with the lattice distortion where 
$3d_{3x^2-r^2}$ and $3d_{3y^2-r^2}$ orbitals alternately align in the $ab$ 
plane. \cite{goodenough,kanamori,matsumoto,rodriguez}
With doping of holes, a variety of spin, charge and orbital ordered phases appears. 
As for the spin degrees of freedom among them, 
its ordering and excitations have been examined experimentally 
in detail by the neutron scattering experiments. 
On the contrary, the charge and orbital orderings 
and excitations still remain to be studied, 
since the experimental probes which directly couple with them  were limited. \cite{ito}
\par
Recently, the charge and orbital orderings in manganites have successfully been observed 
by the resonant x-ray scattering. \cite{murakami1}
Here a Mn $1s$ electron is excited to the Mn $4p$ orbital in the intermediate scattering state 
by tuning the incident x-ray energy to the Mn $K$ edge. 
On resonance, 
the x-ray scattering factors become sensitive dramatically to the charge and orbital states of a Mn $3d$ electron. 
The polarization dependent scattering intensity is utilized to identify the scattering from the orbital ordering. 
Nowadays, this technique is applied to 
several transition metal oxides with charge and orbital orderings. \cite{murakami2,endoh,nakamura,hill,paolasini}
\par
The availability of the third generation synchrotron radiation sources 
promises to detect the charge and orbital excitations 
by the resonant inelastic x-ray scattering (RIXS). \cite{raman,isaccs,hill2,platzman,abbamonte,hasan}
The resonant process provides not only huge enhancement of 
the scattering cross section but also excitations of $3d$ electrons around a Mn ion where x ray is absorbed. 
Thus, this probe is sensitive to the local electronic structures around this ion. 
At the same time, the delocalized character of electrons in a solid is detectable 
by this method, because x ray covers a wide range of the momentum space. 
Actually, momentum dependent RIXS spectra have been recently observed in the insulating cuprates. \cite{tsutsui,hasan} 
These were interpreted as charge excitations from the effective lower Hubbard band to 
the upper Hubbard one across the Mott gap. 
Now this technique is on the point to be applied to the manganites with orbital degree of freedom. 
The Mott gap in the orbital ordered insulating manganites is composed of the upper and lower Hubbard 
bands with different orbital characters unlike the cuprates.
Thus, x ray is scattered from the orbital excitation as well as the charge one in these compounds.   
The x-ray scattering spectrometer of RIXS for manganites 
has recently been constructed on a beamline at the SPring-8. \cite{inami}
Some peak structures are observed in the RXS spectra for LaMnO$_3$ 
around several eV of the energy transfer. 
The polarization and momentum resolved measurements 
will provides a great opportunity to observe excitations in correlated electron systems 
with orbital degeneracy. 
\par
In this paper, we present a theoretical framework of RIXS to study charge and orbital excitations 
in orbital ordered insulating manganites. 
We focus on RIXS from the individual charge and orbital excitations 
caused by the transitions from occupied to unoccupied electronic states across the Mott gap.  
The present theory is applicable to analyze the observed RIXS spectra in LaMnO$_3$ mentioned above. \cite{inami} 
Moreover, an identification of RIXS from the individual excitations 
promises to detect the collective orbital excitation  
proposed theoretically in Ref.~\onlinecite{ishihara1}. 
The scattering cross section is formulated by the Liouville operator method 
where the local electron correlation and itinerant nature of the excitations is treated on an equal footing. 
The cross section is expressed by the charge and orbital correlation functions 
associated with local corrections. 
The RIXS spectra are calculated numerically as functions of momentum, polarization and type of the 
orbital ordered states. 
Through the calculations, we propose that RIXS provides a great opportunity to study 
the electronic excitations in correlated electron systems 
with orbital degeneracy. 
\par
In Sec.~II, a formulation of the scattering cross section  
based on the Liouville operator method is presented.  
The numerical results of the RIXS spectra are shown in Sec.~III. 
Sec.~IV is devoted to the summary and discussion.  
\section{model and formulation}
Let us formulate the scattering cross section of RIXS in insulating manganites with orbital ordering. 
Consider the scattering of x ray with momentum $\vec k_i$ , energy $\omega_i$ 
and polarization $\lambda_i $ to $\vec k_f$, $\omega_f$ and $\lambda_f$.
The electronic states at the initial, intermediate and final states in the scattering process 
are denoted as  
$|i \rangle$, $|m\rangle$ and $|f\rangle$  
with energy $\varepsilon_i$, $\varepsilon_m$ and $\varepsilon_f$, respectively. 
The differential cross section of RIXS is given by \cite{blume,ishihara2}
\begin{equation}
{d^2 \sigma \over d \Omega d \omega_f}= A {\omega_f \over \omega_i}  \sum_{f  }|S|^2 
 \delta(\varepsilon_f+\omega_f-\varepsilon_i-\omega_i) , 
 \label{eq:sigma}
\end{equation}
where 
\begin{eqnarray}
S&=&  \sum_m \Biggl \{ 
{ \langle f | \vec j_{-k_i} \cdot \vec e_{k_i \lambda_i }|m \rangle 
  \langle m | \vec j_{ k_f} \cdot \vec e_{k_f \lambda_f}  |i \rangle 
  \over \varepsilon_i-\varepsilon_m-\omega_f} \nonumber \\
&+&
{\langle f | \vec j_{k_f}  \cdot \vec e_{k_f \lambda_f} | m \rangle 
 \langle m | \vec j_{-k_i} \cdot \vec e_{k_i \lambda_i}  | i \rangle 
 \over \varepsilon_i-\varepsilon_m+\omega_i+i \Gamma}
 \Biggr \} ,
 \label{eq:s}  
\end{eqnarray}
and $A=(e^2/mc^2)^2$.
$ \vec e_{k \lambda }$ is the polarization vector of x ray,   
$\Gamma$ is the damping of a core hole and 
$\vec j_{ k}$ is the current operator defined by 
$ \vec j_{k}=\sum_{l} e^{-i \vec k \cdot \vec r_l} \vec j_{ l}$. 
Since the dipole transition is dominant in RIXS at the Mn$^{3+}$ $K$-edge, \cite{murakami1,murakami2}
this operator is given by 
\begin{equation}
j_{l \alpha}={B \over \sqrt{m}} \sum_\sigma p^\dagger_{l \alpha} s_{l \sigma}+H.c. . 
\label{eq:curr}
\end{equation}
$p^\dagger_{l \alpha}$ and $s_{l \sigma}$ are 
the creation operator of a Mn $4p_{\alpha}$ electron and the annihilation one of a 
Mn $1s$ electron, respectively, at site $l$ with Cartesian coordinate $\alpha$ and spin $\sigma$.  
The coupling constant $B$ in Eq.~(\ref{eq:curr}) between the current and x ray is defined by 
\begin{equation}
B={1 \over 2} \int d \vec r \phi_{4p_\alpha} (\vec r)^\ast (-i \nabla_\alpha) \phi_{1s} (\vec r) , 
\label{eq:bbb}
\end{equation}
with the atomic wave function $\phi_{m}(\vec r)$ $(m=4p,1s)$. 
As shown in Ref.~\onlinecite{ishihara2}, the differential scattering cross section 
is represented by the correlation function of the electronic polarizability 
operator $\alpha_{l \beta \alpha}$ as 
\begin{eqnarray}
{d^2 \sigma \over d \Omega d \omega_f}&=&A {\omega_f \over \omega_i}
\sum_{\alpha \beta \alpha' \beta'}
P_{\beta' \alpha' } P_{ \beta \alpha}
\Pi_{\beta' \alpha' \beta \alpha}(\omega, \vec K) , 
\label{eq:sigma2}
\end{eqnarray}
where $\Pi_{\beta' \alpha' \beta \alpha}(\omega, \vec K)$ is the 
Fourier transform of the correlation function of this operator given by 
\begin{eqnarray}
\Pi_{\beta' \alpha' \beta \alpha}(\omega, \vec K)&=&{1 \over 2 \pi}
\int dt e^{i \omega t} \sum_{ll'} e^{-i\vec K \cdot (\vec r_{l'}-\vec r_{l})}
\nonumber \\
& \times&
\langle i |\alpha_{l' \beta' \alpha'}(t)^\dagger 
\alpha_{l \beta \alpha}(0)| i  \rangle , 
\label{eq:pialal}
\end{eqnarray}
with $\vec K=\vec k_i-\vec k_f$, $\omega=\omega_i-\omega_f$ 
and $P_{\beta \alpha }=(\vec e_{ k_f \lambda_f})_{\beta}   (\vec e_{k_i \lambda_i})_{\alpha}$. 
$\alpha_{l \beta \alpha }(t)$ is the Heisenberg representation of the electronic polarizability 
$ \alpha_{l \beta \alpha }$ at site $l$. 
Now this operator is represented by the Liouville operator $L$ as follows; 
\begin{equation}
\alpha_{l \beta \alpha}= j_{l \beta} {1 \over L-\omega_i+i\Gamma} j_{l \alpha}
                       + j_{l \alpha} {1 \over L^\dagger+\omega_i} j_{l \beta} . 
\label{eq:alpha}
\end{equation}
$L$ is defined by the equation of motion for a Heisenberg operator $O(t)$: \cite{forster}
\begin{equation}
i \partial_t O(t)=[O(t), H]=-L O(t) . 
\label{eq:eqmot}
\end{equation}
On resonance, 
the first term in Eq.~(\ref{eq:alpha}) provides dominant processes in the scattering.  
\par
In the intermediate scattering state of RIXS, 
a Mn $1s$ electron is excited to the Mn $4p$ orbital at a Mn site where x ray is absorbed. 
As discussed in Ref.~\onlinecite{ishihara1}, 
one of the dominant interactions which 
cause charge and orbital excitations of the Mn $3d$ electrons is the 
local Coulomb interactions between $3d$ electrons and the $1s$ hole and/or the $4p$ electron at this site. 
Thus, the RIXS spectra largely depend on the local electronic structure. 
Once the excitations occur at a Mn site, these propagate in a crystal lattice.  
Such itinerant nature of the electronic excitations can be observed by the momentum 
resolved measurements, since the wave length of x ray is comparable to the lattice constant. 
In order to take into account these two characteristics  of RIXS on an equal footing, 
we adopt the following two steps in the formulation of the scattering cross section of RIXS: 
[1] The polarizability operator $\alpha_{l \beta \alpha}$ in Eq.~(\ref{eq:alpha}) is 
expanded by the local operator products at a Mn site $l$. 
Here the Hamiltonian $H_l$ defined at this site is utilized. 
[2] The correlation function 
$\langle i |\alpha_{l' \beta' \alpha'}(t)^\dagger \alpha_{l \beta \alpha}(0)| i  \rangle$ 
in Eq.~(\ref{eq:pialal}) is calculated by the tight-binding Hamiltonian for the Mn 
$3d$ electrons $H_{3d}$ defined in a crystal lattice. 
Contributions from the O $2p$ orbitals which are not included in the present model explicitly 
will be discussed in Sec.~IV. 
\par
Let us introduce the Hamiltonian $H_l$ defined at site $l$ where x ray is absorbed: 
\begin{equation}
H_l=H_l^{(3d)}+H_l^{ (1s,4p)}+H_l^{( 3d-1s,4p)} . 
\label{eq:hamiltonian}
\end{equation}
The first term $H_l^{(3d)}$ represents the Mn $3d$ system  
where two $e_g$ orbitals and a localized spin for $t_{2g}$ electrons are considered. 
This is represented by a sum of the two terms; 
\begin{equation}
H_l^{(3d)}=H_l^\varepsilon+H_l^U , 
\label{eq:h3d}
\end{equation}
where 
\begin{eqnarray}
H_l^\varepsilon=\sum_{\gamma \sigma}  \varepsilon_{d} d_{l \gamma \sigma}^\dagger d_{l \gamma \sigma} , 
\label{eq:3d0}
\end{eqnarray}
and 
\begin{eqnarray}
H_l^U&=&U \sum_{ \gamma} n_{l \gamma \uparrow}n_{l \gamma \downarrow}
     +U'{1 \over 2}\sum_{ \gamma} n_{l \gamma }        n_{l -\gamma }
     \nonumber \\
     &+&J \sum_{ \gamma \sigma \sigma'} 
      d_{l \gamma \sigma }^\dagger  d_{l -\gamma \sigma' }^\dagger
      d_{l \gamma \sigma' }         d_{l -\gamma \sigma }
    -J_H \vec s_{l} \cdot \vec S_{t i} . 
\label{eq:3du}
\end{eqnarray}
$d_{l \gamma \sigma}$ is the annihilation operator of the $3d$ $e_g$ electron at site $l$ 
with orbital $\gamma(=3z^2-r^2, x^2-y^2)$ and spin $\sigma(=\uparrow, \downarrow)$. 
$n_{l \gamma}=\sum_{\sigma} n_{l \gamma \sigma}
             =\sum_{\sigma} d_{l \gamma \sigma}^\dagger d_{l \gamma \sigma} $ 
is the number operator,  
$\vec s_l={1 \over 2}\sum_{s s' \gamma} d_{l \gamma s}^\dagger \vec \sigma_{s s'} d_{l \gamma s'}$ 
is the spin operator for the $e_g$ electrons and 
$\vec S_{tl}$ is the spin operator for the $t_{2g}$ electrons with $S=3/2$.
$-\sigma$ ($-\gamma$) indicates a spin (orbital) which has an opposite direction to $\sigma$ ($\gamma$).   
$U$, $U'$ and $J$ in Eq.~(\ref{eq:3du}) are 
the intra- and inter-orbital Coulomb interactions and the exchange 
interactions, respectively, 
and $J_H$ is the Hund coupling between $e_g$ and $t_{2g}$ spins.
The second term in Eq.~(\ref{eq:hamiltonian}) 
describes the Mn $1s$ and $4p$ electron energy levels: 
\begin{eqnarray}
H_l^{(1s,4p)}=
\varepsilon_s    \sum_{\sigma}        s_{l \sigma}^\dagger        s_{l \sigma}
               +\varepsilon_{p } \sum_{\alpha \sigma} p_{l \alpha \sigma}^\dagger p_{l \alpha \sigma} , 
\label{eq:h1s4p}
\end{eqnarray}
and the third term gives the Coulomb interactions between the Mn $1s$, $3d$ and $4p$ electrons: 
\begin{eqnarray}
H_l^{(3d-1s,4p)}&=&n_l^s \Bigl (V_{sd}  n_l+ V_{sp} n_l^p \Bigr ) 
\nonumber \\
            &+&\sum_{\gamma \alpha} n_{l \alpha}^p 
                              \Bigl( V_{dp}^{\gamma, \alpha} n_{l \gamma} 
                                    +W_{dp}^{\gamma, \alpha} m_{l \gamma} \Bigr ) , 
\label{eq:h3d1s4p}
\end{eqnarray}
whre 
$n_l^{s}=\sum_{\sigma}s_{l \sigma}^\dagger s_{l \sigma}$ and 
$n_l^p=\sum_{\alpha} n_{l \alpha}^p=\sum_{\alpha \sigma} p_{l \alpha \sigma }^\dagger p_{i \alpha \sigma}$.  
The first two terms in Eq.~(\ref{eq:h3d1s4p})
correspond to the core hole potentials and   
the second two describe the Coulomb interactions between the Mn $3d$ and $4p$ electrons.
The latter involve the operator $m_{l \gamma}=\sum_{\sigma} d_{l -\gamma \sigma}^\dagger d_{l \gamma \sigma}$ 
which describes the orbital fluctuation for $3d$ electrons. 
Explicit forms of the diagonal and off-diagonal components of 
the Coulomb interactions $V_{dp}^{\gamma, \alpha}$ and $W_{dp}^{\gamma, \alpha}$ 
are given as
\begin{equation} 
V_{dp}^{\gamma, \alpha}=F_0+4 F_2 \cos \Bigl(\theta_\gamma-{2 \pi \over 3}m_\alpha \Bigr) , 
\end{equation}
and 
\begin{equation}
W_{dp}^{\gamma, \alpha}=4F_2 \sin \Bigl(\theta_\gamma-{2 \pi \over 3}m_\alpha \Bigr) , 
\label{eq:www}
\end{equation} 
respectively, with $(m_x,m_y,m_z)=(1,2,3)$. 
$F_{0(2)}$ is the Slater integral between $3d$ and $4p$ electrons 
and an angle $\theta_\gamma$ specifies the occupied $3d$ orbital as 
\begin{equation}
|\theta_\gamma \rangle=\cos({\theta_\gamma \over 2})|d_{3z^2-r^2}\rangle
                          -\sin({\theta_\gamma \over 2})|d_{x^2-y^2}\rangle . 
\label{eq:theta}
\end{equation}
\par
Being based on $H_l$ given in Eq.~(\ref{eq:hamiltonian}), 
$\alpha_{l \beta \alpha}$ is expanded by the local operator products at site $l$. 
The Liouville operator $L$ is divided into the 
unperturbed and perturbed parts, i.e. 
$L=L_0+L'$. $L_0$ and $L'$ correspond to the Hamiltonian 
$H_l^0=H_l^\varepsilon+H_l^{(1s,4p)}+\langle H_l^U \rangle _H+ \langle H_l^{(3d-1s,4p)} \rangle_{H}$
and 
$H_l'=\Delta (H_l^U)+\Delta(H_l^{(3d-1s,4p)} )$, respectively, 
where $\hat O=\langle \hat O \rangle_{H}+\Delta (\hat O)$ for an operator $\hat O$ 
and $\langle \hat O \rangle_{H}$ is its Hartree part. 
Up to the first order of $L'$, 
$\alpha_{l \beta \alpha}$ is expressed as follows, 
\begin{eqnarray}
\alpha_{l \beta \alpha}&=& 
{|B|^2 \over m} \sum_{\sigma_1 \sigma_2} 
J_{l \beta \sigma_2}^\dagger
\Biggl (
{1 \over L_0-\omega_i+i\Gamma} 
\nonumber \\
&-&
{1 \over L_0-\omega_i+i\Gamma}L' {1 \over L_0-\omega_i+i\Gamma} 
\Biggr )
J_{l \alpha \sigma_1} , 
\end{eqnarray}
with $J_{l \alpha \sigma}=p_{l \alpha \sigma}^\dagger s_{l \sigma}$. 
The first and second terms correspond to the elastic and inelastic scatterings, respectively. 
The second term is calculated  
by utilizing the equations of motion of the operators and is given by 
\begin{equation}
\alpha_{l \beta \alpha}= 
-{|B|^2 \over m} \delta_{\alpha \beta} \sum_{\sigma_1 \sigma_2} 
J_{l \alpha \sigma_2}^\dagger J_{l \alpha \sigma_1}
\sum_{m=1,2} \sum_{\sigma \gamma} 
O_{l \gamma \sigma}^{(m)} D_{ \gamma \sigma \alpha}^{(m)} , 
\label{eq:alpfin}
\end{equation}
with 
\begin{equation}
 D_{ \gamma \sigma \alpha }^{(m)}
={1 \over E_{ \gamma \sigma \alpha }^{(m)}-\omega_i+i \Gamma}
C^{(m)}_{\gamma \alpha}
{1 \over E_{ \alpha }-\omega_i+i \Gamma} . 
\label{eq:ddd}
\end{equation}
$m=(1,2)$ describes 
the intra- ($m=1$) and inter-orbital ($m=2$) excitations for the $3d$ electrons. 
The operator $O_{l \gamma \sigma}^{(m)}$ represents these excitations and 
$C^{(m)}_{\gamma \alpha}$ and 
$E_{\gamma \sigma \alpha}^{(m)}$ are their amplitudes and excitation energies, respectively. 
The explicit forms of them are given as follows,  
\begin{equation}
O_{l \gamma \sigma}^{(1)}=\delta n_{l \gamma \sigma}, \ \ \  
O_{l \gamma \sigma}^{(2)}=\delta m_{l \gamma \sigma}^\dagger ,   
\label{eq:ope}
\end{equation}
\begin{equation}
C^{(1)}_{\gamma \alpha}=V_{sd}-V_{dp}^{\gamma, \alpha}, \ \ \ 
C^{(2)}_{\gamma \alpha}=-W_{dp}^{\gamma, \alpha} , 
\label{eq:amplitude}
\end{equation}
\begin{equation}
E_{ \gamma \sigma \alpha}^{(1)}=-E_{\alpha } , \ \ \ 
E_{ \gamma \sigma \alpha}^{(2)}=
-(\varepsilon_{-\gamma \sigma}-\varepsilon_{\gamma \sigma}+E_{\alpha})  , 
\label{eq:inte}
\end{equation}
where
$\hat O=\langle \hat O \rangle+\delta \hat O$ for an operator $\hat O$ and 
\begin{equation}
E_{\alpha }=\varepsilon_s-\varepsilon_{p}
+V_{sd}\langle n_{l} \rangle
-\sum_{\gamma} V_{dp}^{\gamma \alpha}-V_{sp} , 
\end{equation}
\begin{eqnarray}
\varepsilon_{\gamma \sigma}&=&
\varepsilon_{d }
+U\langle n_{l \gamma -\sigma} \rangle
+U' \langle n_{l -\gamma} \rangle
-J \langle n_{l -\gamma \sigma} \rangle
\nonumber \\
&-&J_H {1 \over 2} \varepsilon_{\sigma} \langle S_{l t}^z \rangle
+V_{sd}\langle n_{l}^s \rangle
+\sum_{\alpha}V_{dp}^{\gamma \alpha} \langle n_{l \alpha}^p \rangle , 
\end{eqnarray}
with $(\varepsilon_\uparrow, \varepsilon_\downarrow)=(1,-1)$. 
Then, Eq.~(\ref{eq:alpfin}) denotes the following inelastic scattering processes:  
the incident x ray excites an electron from the $1s$ orbital to the $4p$ one 
with the excitation energy $E_\alpha$ at a Mn site $l$ where x ray is absorbed. 
In the intermediate scattering state, 
charge and orbital excitations represented by $O_{l \gamma \sigma}^{(m)}$
occur with amplitudes $C^{(m)}_{\gamma \alpha}$ at this Mn site. 
Finally, the $4p$ electron returns back to the $1s$ orbital with emitting the second x ray. 
The final form of the correlation function of the polarizability is given by 
\begin{eqnarray}
\Pi_{\beta' \alpha' \beta \alpha}(\omega, \vec K)&=&{1 \over 2 \pi}
\int dt e^{i \omega t} \sum_{ll'} e^{-i\vec K \cdot (\vec r_{l'}-\vec r_{l})}
\nonumber \\
&\times&
\Pi_{\beta' \alpha' \beta \alpha}(t,\vec r_{l'}-\vec r_{l}) , 
\label{eq:corrr0}
\end{eqnarray}
with 
\begin{eqnarray}
\Pi_{\beta' \alpha' \beta \alpha}(t,\vec r_{l'}-\vec r_{l})&=&
{|B|^4 \over m^2} \delta_{\beta' \alpha'}\delta_{\beta \alpha}
\sum_{mm'}\sum_{\sigma \sigma' \gamma \gamma'}
\nonumber \\
&\times&
D_{\gamma' \sigma' \alpha'}^{(m')\ast} D_{\gamma \sigma \alpha}^{(m)} 
\langle O_{l' \gamma' \sigma'}^{(m')}(t)^\dagger O_{l \gamma \sigma}^{(m)}(0) \rangle . 
\label{eq:corrr}
\end{eqnarray}
Here it is assumed that 
$ J_{l \alpha \sigma_2}^\dagger J_{l \alpha \sigma_1}=\delta_{\sigma_1 \sigma_2}$ and 
$\langle i | \cdots| i \rangle$ is replaced by 
the thermal average $\langle \cdots \rangle$. 
We note that the scattering cross section is expressed by 
the correlation functions of the charge and orbital fluctuations of the Mn $3d$ electrons associated with the local 
corrections  $D_{\gamma \sigma \alpha}^{(m)}$. 
\par
The correlation function 
$\langle O_{l' \gamma' \sigma'}^{(m')}(t)^\dagger O_{l \gamma \sigma}^{(m)}(0) \rangle$ in Eq.~(\ref{eq:corrr})
is calculated by the tight-binding Hamiltonian $H_{3d}$ for the  Mn $3d$ electrons defined in a crystal lattice: 
\begin{equation}
H_{3d}=\sum_l H_l^{(3d)}+H_t , 
\label{eq:h3d2}
\end{equation}
where
$ H_l^{(3d)}$ is given in Eq.~(\ref{eq:h3d}).
The second term represents the electron hopping between nearest neighboring Mn sites  
$l$ and $l'$ with orbitals $\gamma$ and $\gamma'$, respectively.; 
\begin{equation}
H_t=\sum_{\langle ll' \rangle \gamma \gamma' \sigma} 
t_{ll'}^{\gamma \gamma'} 
d_{l \gamma \sigma}^\dagger d_{l' \gamma' \sigma} +H.c. , 
\label{eq:ht}
\end{equation}
where $t_{ll'}^{\gamma \gamma'}$ is the hopping integral and its orbital dependence 
is determined by the Slater-Koster parameters. \cite{slater}
\par
We apply the present formulae of the scattering cross section to LaMnO$_3$. 
A simple cubic lattice consisting of Mn$^{3+}$ is considered.  
The $A$-type antiferromagnetic structure and 
the $C$-type orbital ordered one, where $d_{3x^2-r^2}$ and $d_{3y^2-r^2}$ 
orbitals alternately align in the $ab$ plane, 
is introduced by considering the Jahn-Teller type lattice distortion observed in LaMnO$_3$. 
The mean field approximation is adopted in $H_{3d}$; 
$\langle s_k^z \rangle={1 \over 4}\langle S_{tk}^z \rangle={1 \over 2}\delta_{k=(00\pi)}$, 
$\langle n_{k 3z^-2r^2} \rangle={1 \over 2}\delta_{k=(000)}$, 
$\langle n_{k  x^2-y^2} \rangle={3 \over 2}\delta_{k=(000)}$ and  
$\langle m_{k 3z^2-r^2} \rangle=\langle m_{k  x^2-y^2} \rangle={1 \over 4}\delta_{k=(\pi \pi 0)}$ 
where $\hat O_k$ is the Fourier transform of an operator $\hat O_l$. 
As for the lattice degree of freedom, 
the adiabatic approximation is introduced in the intermediate and final scattering states 
where the lattice distortions are assumed  to be unchanged from those in the initial scattering state. 
Contributions from the Jahn-Teller coupling to the excitation energies are 
interpreted to be included implicitly in the inter-orbital Coulomb interaction $U'$. 
%
%
\begin{figure}
\epsfxsize=0.8\columnwidth
\centerline{\epsffile{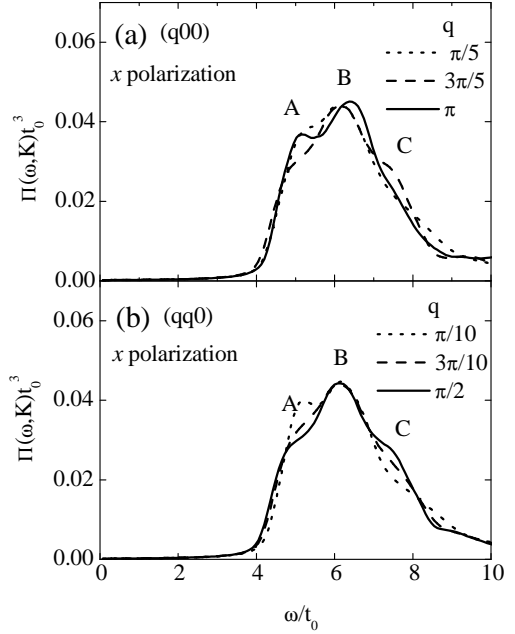}}
\caption{
The RIXS spectra $\Pi_{\beta' \alpha' \beta \alpha}(\omega, \vec K)$ for 
the $(d_{3x^2-r^2},d_{3y^2-r^2})$-type orbital ordered state. 
The momentum transfer is (a) $\vec K=(q 0 0)$ and 
(b) $\vec K=(q q 0)$ in the cubic Brillouin zone. 
The polarization of x ray is chosen to be $\alpha=\alpha'=\beta=\beta'=x$ (the $x$ polarization).  
The energy parameters are chosen to be  
$U=9$, $U'=7$, $J=J_H=1$, $V_{sd}=9.5$, $F_0=7$, $F_2=0.35$ and $\Gamma=1.5$ as a unit of $t_0$. 
$t_0$ is estimated to be about 0.5$\sim$ 0.7eV. 
}
\label{fig:fig1}
\end{figure}
\section{numerical results}
The calculated results of the RIXS spectra are presented in Fig.~1. 
$\Pi_{\beta' \alpha' \beta \alpha}(\omega, \vec K)$ defined in Eqs.~(\ref{eq:corrr0}) and (\ref{eq:corrr})
is plotted as a function of the energy transfer $\omega$ of x ray. 
The elastic component located at $\omega=0$ is not shown. 
The momentum transfer $\vec K$ is chosen to be $(q00)$ for Fig.~1(a) 
and $(qq0)$ for Fig.~1(b) where the notation in a simple cubic lattice is used. 
The polarization of x ray is assumed to be parallel to the $x$ direction, i.e. $\alpha=\alpha'=\beta=\beta'=x$. 
The energy parameters are chosen to be 
$U=9$, $U'=7$, $J=J_H=1$, $V_{sd}=9.5$, $F_0=7$, $F_2=0.35$ and $\Gamma=1.5$ in units of $t_0$ 
which is the hopping integral between $d_{3z^2-r^2}$ orbitals in the $z$ direction.  
$t_0$ is estimated to be about 0.5$\sim$ 0.7eV. 
The energy of the incident x ray is redefined as 
${\tilde \omega_i}=\omega_i-(\varepsilon_p-\varepsilon_s-V_{sp}-V_{sd})$ in  
Eq.~(\ref{eq:inte}). 
We assume that the smallest energy among 
$E_{ \gamma \sigma \alpha }^{(m)}$ and $E_{ \alpha }$ in Eq.~(\ref{eq:ddd}) 
corresponds to the excitation energy at the $K$ edge and $\tilde \omega_i$ is fixed at this energy. 
%
%
\begin{figure}
\epsfxsize=0.8\columnwidth
\centerline{\epsffile{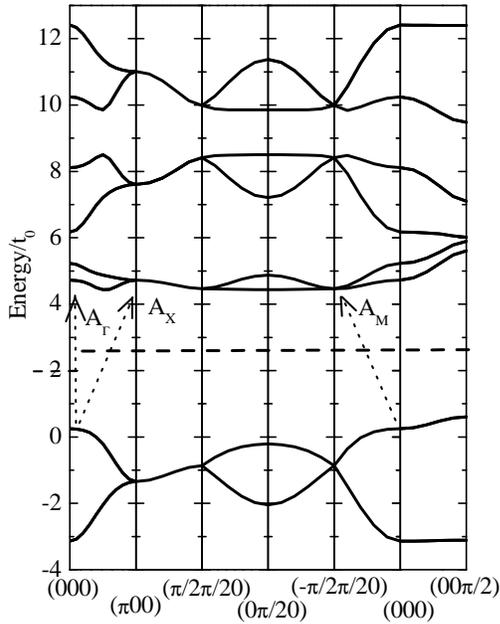}}
\caption{
The electron energy bands for the $(d_{3x^2-r^2},d_{3y^2-r^2})$-type orbital ordered state. 
The broken line represents the chemical potential located at the center of the occupied and lowest unoccupied bands. 
The parameter values are the same with those in Fig.~1.
The origin of the vertical axis is arbitrary. 
The dotted arrows indicated by $A_{\Gamma}$, $A_{X}$ and $A_{M}$ 
represent the main contributions to the peak $A$ in the RIXS spectra (see Fig.~1) around 
$\vec K=(000)$, $(\pi 00)$ and $({\pi \over 2} {\pi \over 2} 0)$, respectively. 
}
\label{fig:fig2}
\end{figure}
The RIXS spectra shown in Fig.~1 do not have an intensity up to about 2$t_0$ 
which corresponds to the Mott gap. 
That is, the spectra are attributed to the individual electronic excitations 
from occupied to unoccupied electronic states across the Mott gap. 
The remarkable momentum dependence is seen in the RIXS spectra; 
near the $(000)$ point, two peaks and one shoulder are observed at about 
$5t_0$ and $6t_0$ and $7.5t_0$, respectively. 
These are denoted as $A$, $B$ and $C$ in the figure.
The lowest peak structure becomes remarkable 
with increasing $\vec K$ along the $<h00>$ direction. 
On the other hand,  
this structure is smeared out near the $({\pi \over 2} {\pi \over 2} 0)$ point,
\par
%
%
\begin{figure}
\epsfxsize=0.8\columnwidth
\centerline{\epsffile{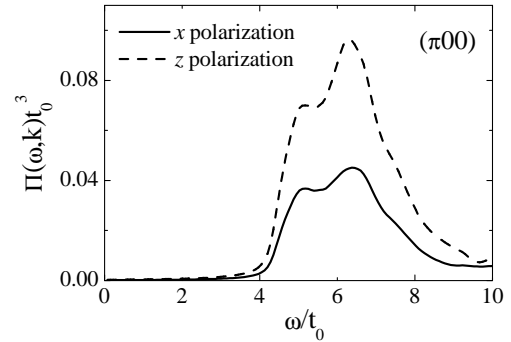}}
\caption{
The polarization dependence of the RIXS spectra $\Pi_{\beta' \alpha' \beta \alpha}(\omega, \vec K)$ for 
the $(d_{3x^2-r^2},d_{3y^2-r^2})$-type orbital ordered state. 
The momentum transfer is $\vec K=(\pi 0 0)$. 
The solid and broken lines represent the spectra 
for $\alpha=\alpha'=\beta=\beta'=x$ (the $x$ polarization) and 
$\alpha=\alpha'=\beta=\beta'=z$ (the $z$ polarization), respectively.  
Other parameter values are the same with those in Fig.~1. 
}
\label{fig:fig3}
\end{figure}
The energy and momentum dependence of the RIXS spectra is well explained 
by the Hartree-Fock band structure presented in Fig.~2. 
The eight bands each of which are doubly degenerate are attributed to the four Mn ions 
with spin and orbital degrees of freedom in a unit cell.  
The occupied bands have the $d_{3x^2-r^2}$ and $d_{3y^2-r^2}$ orbital characters, which are termed the majority 
orbitals hereafter, 
and the lowest unoccupied bands have the $d_{y^2-z^2}$ and $d_{z^2-x^2}$ orbital ones (the minority orbitals). 
The Mott gap opens between the two and is of the order of $U'-J$ in the limit of $U' >> t_0$.  
The unoccupied bands with higher energies correspond to the minority spin bands.  
The dispersion relations of all bands are weak in the $z$ direction, because the electron hopping 
is suppressed in this direction due to the $A$ type antiferromagnetic structure. 
It is interpreted that the main RIXS spectra in Fig.~1 are attributed to the transition from 
the occupied band (the lower Hubbard band) to the lowest unoccupied band (the upper Hubbard band). 
Weak spectral weights in the regions of $8.5<\omega/t_0$ originate in the excitations to the higher bands. 
These transitions are almost prohibited, 
since the operators $O_{\l \gamma \sigma}^{(m)}$ 
in Eqs.~(\ref{eq:ope}) do not change the spin states of the $3d$ electrons. 
It is noted that the lowest unoccupied bands show the almost flat dispersion relation, 
because the electron hopping between the $d_{y^2-z^2}$ and $d_{z^2-x^2}$ orbitals is forbidden in the $xy$ plane. 
Thus, the global shape of the main spectra reflects from the density of states of 
the occupied band.  
The dominant contributions to the peak $A$ around $\vec K=(000)$, $(\pi00)$, and 
$({\pi \over 2}{\pi \over 2}0)$ in Fig.~1 
are attributed to the transitions denoted as $A_\Gamma$, $A_X$ and $A_M$ in Fig.~2, respectively.
It is noted that the difference between the transitions $A_M$ and $A_X$ 
is caused by the curvature of the unoccupied band; 
$\nabla_{x,y} E^{(u)}(\vec k)<0$ around $\vec k=(\pi 0 0)$ and 
$\nabla_{x,y} E^{(u)}(\vec k)>0$ around $\vec k=({\pi \over 2}{\pi \over 2}0)$ 
where $E^{(u)}(k)$ is the energy of the lowest unoccupied band 
and $\nabla_{x,y}$ is a differential operator 
in the ($k_x$, $k_y$) plane. 
As for the occupied band, we find that $\nabla_{x,y} E^{(o)}(\vec k)<0$ around $\vec k=(000)$
where $E^{(o)}(\vec k)$ is the energy of the occupied band. 
Thus, the van Hove singularity, where 
the condition $\nabla_{x,y}E^{(u)}(\vec k)-\nabla_{x,y}E^{(o)}(\vec k')=0$ is satisfied, 
exists in a wide region of the momentum space for the $A_X$ transition. 
The difference between 
$\nabla_{x,y} E^{(u)}(\vec k)$'s  around $\vec k=(\pi 0 0)$ and 
$({\pi \over 2}{\pi \over 2}0)$
is attributed to the fact that 
hopping integrals between the occupied orbitals ($d_{3x^2-r^2}$, $d_{3y^2-r^2}$) 
and the unoccupied ones ($d_{y^2-z^2}$, $d_{z^2-x^2}$) disappear at the $({\pi \over 2}{\pi \over 2}0)$ point.  
Consequently, the peak $A$ becomes remarkable around $\vec K=(\pi00)$ as seen in Fig.~1. 
\par
%
\begin{figure}
\epsfxsize=0.8\columnwidth
\centerline{\epsffile{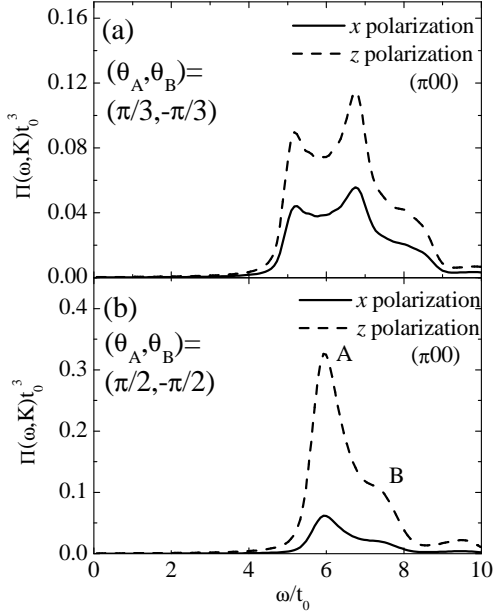}}
\caption{
The RIXS spectra $\Pi_{\beta' \alpha' \beta \alpha}(\omega, \vec K)$ for 
(a) the $(d_{x^2-z^2},d_{y^2-z^2})$-type ($(\theta_A,\theta_B)=(\pi/3,-\pi/3)$) and (b) the 
${1 \over \sqrt{2}}(d_{3x^2-r^2}+d_{x^2-z^2},d_{3y^2-r^2}+d_{y^2-z^2})$-type ($(\theta_A,\theta_B)=(\pi/2,-\pi/2)$)
orbital ordered states. 
The momentum transfer is $\vec K=(\pi 0 0)$. 
The solid and broken lines represent the spectra 
for $\alpha=\alpha'=\beta=\beta'=x$ (the $x$ polarization) and 
$\alpha=\alpha'=\beta=\beta'=z$ (the $z$ polarization), respectively.  
Other parameter values are the same with those in Fig.~1. 
}
\label{fig:fig4}
\end{figure}
The polarization dependence of the RIXS spectra is shown in Fig.~3. 
The polarization of x ray is chosen to be 
parallel to the $x$ direction ($\alpha=\alpha'=\beta=\beta'=x$) for Fig.~3(a) 
and the $z$ direction ($\alpha=\alpha'=\beta=\beta'=z$) for Fig.~3(b). 
The momentum transfer is fixed at $\vec K=(\pi 0 0 )$ in both cases. 
The spectrum is much enhanced for the $z$ polarization,  
although the global shape of the spectrum is insensitive to the polarization.  
This result arises from 
the factor $C_{\gamma \alpha}^{(m)}$ given in Eq.~(\ref{eq:amplitude});  
$C_{\gamma \alpha}^{(m=2)}(=-W_{dp}^{\gamma, \alpha})$ 
is an amplitude of the excitations from the occupied $3d$ orbital $\gamma$ 
to the unoccupied orbital $-\gamma$,
when the $4p_\alpha$ orbital is occupied by an electron. 
Consider the two kinds of Mn sites where $d_{3x^2-r^2}$ and $d_{3y^2-r^2}$ orbitals are occupied. 
In the case where a $1s$ electron is excited to the $4p_\alpha$ orbital by x ray, 
the amplitudes $(W_{dp}^{3x^2-r^2, \alpha},W_{dp}^{3y^2-r^2, \alpha})$ are given by   
$(0, 2 \sqrt{3} F_2)$ for $\alpha=x$ and 
$( 2\sqrt{3}  F_2,-2 \sqrt{3}  F_2)$ for $\alpha=z$. 
We note that $W_{dp}^{3x^2-r^2, x}=0$ because of the symmetry of the wave functions. 
This is clearly shown by the explicit form of this interaction: 
\begin{eqnarray}
W_{dp}^{3x^2-r^2, x}&=&
\int d\vec r d \vec r' \phi_{4p_x}^\ast(\vec r) \phi_{4p_x}(\vec r) v(|\vec r-\vec r'|)
\nonumber \\
&\times&
\phi_{3d_{y^2-z^2}}^\ast(\vec r') \phi_{3d_{3x^2-r^2}}(\vec r') , 
\end{eqnarray}
where $\phi_\gamma(\vec r)$ is the atomic wave function of the orbital $\gamma$ 
and $v(|\vec r|)$ is the Coulomb interaction. 
The integrand changes its sign under replacement of $y$ by $z$. 
Thus, the orbital excitation does not occur at the $d_{3x^2-r^2}$ site  
by x ray with the $x$ polarization.  
\par
%

%
\begin{figure}
\epsfxsize=0.8\columnwidth
\centerline{\epsffile{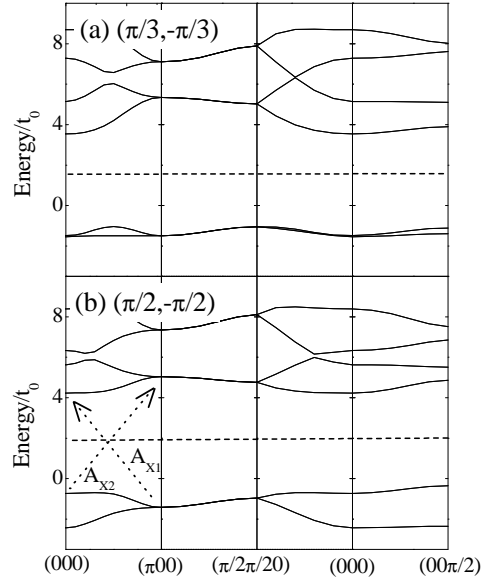}}
\caption{
The electron energy bands for 
(a) the $(d_{x^2-z^2},d_{y^2-z^2})$-type  ($(\theta_A,\theta_B)=(\pi/3,-\pi/3)$) 
and 
(b) the ${1 \over \sqrt{2}}(d_{3x^2-r^2}+d_{x^2-z^2},d_{3y^2-r^2}+d_{y^2-z^2})$-type 
($(\theta_A,\theta_B)=(\pi/2,-\pi/2)$)
orbital ordered states.  
The broken line represents the chemical potential located at the center of the occupied and lowest unoccupied bands. 
The parameter values are the same with those in Fig.~1.
The origin of the vertical axis is arbitrary. 
The dotted arrows in (b) indicated by $A_{X1}$ and $A_{X2}$ 
represent the main contributions to the peak $A$ in the RIXS spectra (see Fig.~4(b)) around 
$\vec K=(\pi 00)$. 
The origin of the vertical axis is arbitrary. 
}
\label{fig:fig5}
\end{figure}
RIXS spectra calculated in several-types of the orbital ordered states are compared in Fig.~4. 
We consider the $C$-type orbital ordered state where the two orbital sublattices exist. 
The occupied orbitals are denoted by the angles $(\theta_A,\theta_B)$ defined in Eq.~(\ref{eq:theta}). 
These values are chosen to be 
$(\theta_A,\theta_B)=(\pi/3,-\pi/3)$ for Fig.~4(a) and $(\pi/2,-\pi/2)$ for Fig.~4(b)  
which correspond to the $(d_{z^2-x^2},d_{y^2-z^2})$ and 
$({1 \over \sqrt{2}}[d_{3x^2-r^2}+d_{z^2-x^2}],{1 \over \sqrt{2}}[d_{3y^2-r^2}+d_{y^2-z^2}])$ 
orbital ordered states, respectively. 
The momentum transfer is fixed at $\vec K=(\pi00)$. 
The Hartree-Fock energy bands in these orbital states are also presented in Fig.~5. 
We note that the energy and polarization dependence of the spectra in Fig.~4(a) 
is similar to that for the $(2\pi/3,-2\pi/3)$ state presented in Fig.~1. 
This is because the occupied and lowest unoccupied bands in the state have 
the $(d_{z^2-x^2},d_{y^2-z^2})$ and $(d_{3x^2-r^2},d_{3y^2-r^2})$ orbital characters, respectively, 
which are opposite to those in the $(2\pi/3,-2\pi/3)$ orbital ordered state, as seen in Figs.~1 and 5(a). 
On the contrary, 
RIXS spectra for the $(\pi/2,-\pi/2)$ state show 
mainly two peak structures located around $\omega=6t_0$ and $7.5t_0$. 
These peaks are indicated as $A$ and $B$ in Fig.~4(b). 
It is worth noting that 
the dispersion relations of the occupied and lowest unoccupied bands in Fig.~5(b) 
are almost symmetrical with 
respect to the center of the Mott gap because of the symmetry of this orbital ordered state. 
Thus, the excitation energies for some transitions, for example, the transitions 
$A_{X1}$ and $A_{X2}$, are nearly degenerate 
unlike the $(\pi/3,-\pi/3)$ and $(2\pi/3,-2\pi/3)$ orbital ordered states. 
The spectra in this orbital ordered state is more sensitive to the polarization of x ray 
than those in the $(\pi/3,-\pi/3)$ and $(2\pi/3,-2\pi/3)$ states. 
This fact arises from the larger anisotropy of the off-diagonal Coulomb interaction, 
i.e., $(W_{dp}^{\pi/2, \alpha},W_{dp}^{-\pi/2, \alpha})=
(2 F_2 ,-2 F_2)$ for the $x$ polarization and 
$( 4  F_2 ,4 F_2)$ for the $z$ polarization 
where $W_{dp}^{\pm \pi/2, \alpha}$ is the off-diagonal Coulomb interaction $W_{dp}^{\gamma, \alpha}$ for 
the $3d$ orbital of $\theta=\pm \pi/2$. 
\section{summary and discussion}
We present, in this paper, a theoretical framework of RIXS from charge and orbital excitations in 
insulating manganites with orbital ordering. 
We formulate the scattering cross section by the Liouville operator method 
from the stand point that both the local and itinerant nature of the excitations 
reflects on the RIXS spectra. 
Then, the cross section is expressed by the correlation functions 
of the charge and orbital excitations associated with the local corrections. 
The general formulae for the scattering cross section is applied to LaMnO$_3$ 
where the $A$-type antiferromagnetic order and the $C$-type orbital order 
of the $(3d_{3x^2-r^2},3d_{3y^2-r^2})$-type are realized. 
The calculated spectra from the individual excitations 
are interpreted by the electronic transition from the occupied to lowest unoccupied 
states across the Mott gap. 
However, the spectra are not the simple joint density of states itself; 
the local correlation effects dominate the unique polarization and orbital dependence of the spectra. 
\par
Let us discuss the relevance of the present results to the experimental ones in manganites. 
LaMnO$_3$ is a charge transfer type insulator where the insulating gap is formed between  
the unoccupied Mn $e_g$ band (the upper Hubbard band) and the O $2p$ band with the $e_g$ symmetry 
being strongly hybridized with the Mn $3d$ ones. 
A value of this gap $\Delta$ is estimated to be about 2$\sim$3eV. \cite{saitoh,arima}
Below this highest occupied band, 
the so-called non-bonding O $2p$ bands exist.  
In the present model, 
where the oxygen ions are not explicitly taken into account, 
the lower Hubbard band is interpreted to be this highest occupied band 
with the strong hybridization between O $2p$ and Mn $3d$ orbitals. 
The calculated RIXS spectra are attributed to the transition from 
this band to the upper Hubbard one and its energy transfer is about $\Delta$. 
In the region of higher energy transfer, 
RIXS spectra arising from the transition from the non-bonding O $2p$ bands to the upper Hubbard ones 
are expected to appear, as is observed at about 6eV in insulating cuprates. \cite{abbamonte,hill}
Below $\Delta$, on the other hand, 
RIXS from the collective orbital excitation termed orbital wave \cite{ishihara0,ishihara1}
is expected to occur. 
The energy of this excitation is characterized by the inter-site interaction between orbitals.
This interaction is of the order of 0.1eV. 
An absolute value of the scattering intensity for the calculated RIXS spectra 
is roughly estimated to be of the order of 10$^{-24}$cm$^{2}$ 
where we assume $B=2\AA^{-1}$ in Eq.~(\ref{eq:bbb}) and $t_0=0.7eV$. 
This is about 50$\sim$100 times larger than the scattering cross section for the Thomson scattering 
$(e^2/mc^2)^2$ which implies an intensity of 1count/sec/1eV for the conventional experimental 
arrangement. \cite{raman} 
It is enough to detect in the present state of the experiments, although the estimation is semi-quantitative.  
With doping of holes into LaMnO$_3$, 
a finite intensity of the spectra may appear inside the gap of the spectra. 
At the same time, a change of the types of the orbital ordered state is expected to occur \cite{endoh,hirota}
and is to be detected by RIXS as shown in the present theory. 
\begin{acknowledgments}
Authors would like to thank 
Y.~Endoh, J.~Mizuki, T.~Inami, and Y.~Murakami 
for their valuable discussions.  
This work was supported by the Grant in Aid from Ministry of Education, 
Science and Culture of Japan, CREST and NEDO. 
Part of the numerical calculation was performed in the HITACS-3800/380 
superconputing facilities in IMR, Tohoku University. 
\end{acknowledgments}

\begin{references}
%
%
\bibitem{imada}
See, for example, 
M.~Imada, A.~Fujimori, and Y.~Tokura,  
Rev.~Mod.~Phys. {\bf 70} 1039 (2000). 
%
\bibitem{tokura}
Y.~Tokura, and N.~Nagaosa, Science {\bf 288}, 462 (2000). 
%
\bibitem{wollan}
E.~O.~Wollan, and W.~C.~Koehler, 
Phys.~Rev. {\bf 100}, 545 (1955). 
%
\bibitem{goodenough}
J. B. Goodenough, 
Phys. Rev. {\bf 100}, 564 (1955).
%
\bibitem{kanamori}
J. Kanamori, J. Phys. Chem. Sol. {\bf 10}, 87 (1959).
%
\bibitem{matsumoto}
G. Matsumoto,  
J. Phys. Soc. Jpn. {\bf 29}, 606 (1970).
%
\bibitem{rodriguez}
J.~Rodriguez-Carvajal, M.~Hennion, F.~Moussa, and A.~H.~Moudden, 
Phys. Rev. B {\bf 57}, R3189 (1998). 
%
\bibitem{ito}
Y.~Ito, and J.~Akimitsu,  J.~Phys.~Soc.~Jpn. {\bf 40}, 1333 (1976).
%
\bibitem{murakami1}
Y. Murakami, H. Kawada, H. Kawata, M. Tanaka, T. Arima, H. Moritomo, and Y. Tokura, 
Phys. Rev. Lett. {\bf 80}, 1932 (1998).
%
\bibitem{murakami2}
Y. Murakami, J. P. Hill, D. Gibbs, M. Blume, I. Koyama, M. Tanaka, 
H. Kawata, T. Arima, T. Tokura, K. Hirota, and Y. Endoh, Phys. Rev. Lett. {\bf 81}, 582 (1998).
%
\bibitem{endoh}
Y. Endoh, K. Hirota, S. Ishihara, S. Okamoto, Y. Murakami
A. Nishizawa, T. Fukuda, H. Kimura, 
N. Nojiri, K. Kaneko, and S. Maekawa, 
Phys. Rev. Lett. {\bf 82}, 4328 (1999). 
%
\bibitem{nakamura}
K. Nakamura, T. Arima, A. Nakazawa, Y. Wakabayashi, and Y. Murakami , 
Phys. Rev. B {\bf 60}, 2425 (1999).
%
\bibitem{hill}
M.~v.~Zimmermann, J.~P.~Hill, D.~Gibbs, M.~Blume, D.~Casa, B.~Keimer, Y.~Murakami, Y.~Tomioka, and Y.~Tokura, 
Phy. Rev. Lett.  {\bf 83}, 4872 (1999).  
%
\bibitem{paolasini}
L.~Paolasini, C.~Vettier, F.~de~Bergevin, F.~Yakhou, D.~Mannix, A.~Stunault, and W.~Neubeck, 
Phy. Rev. Lett.  {\bf 82} (1999) 4719. 
%
\bibitem{raman}
For a review see  
{\it Raman emission by x-ray scattering}
edited by D.~L.~Ederer, and J.~H.~McGuire, 
(World Scientific, Singapore, 1996).
%
\bibitem{isaccs}
E.~D.~Isaccs, P.~M.~Platzman, P.~Metcalf, and J.~M.~Honig, 
Phy. Rev. Lett. {\bf 76}, 4211 (1996).
%
\bibitem{hill2}
J.~P.~Hill, C.~-C.~Kao, W.~A.~L.~Calieve, M.~Matsubara, A.~Kotani, J.~L.~Peng, and R.~L.~Greene,  
Phy. Rev. Lett. {\bf 80}, 4976 (1998).
%
%
\bibitem{platzman}
P.~M.~Platzman, and E.~D.~Isaacs,   
Phy. Rev. B  {\bf 57}, 11107 (1998).
%
\bibitem{abbamonte}
P.~Abbamonte, C.~A.~Burns, E.~D.~Isaacs, P.~M.~Platzman, L.~L.~Miler, S.~W.~Cheong, and M.~V.~Klein,  
Phy. Rev. Lett. {\bf 83}, 860 (1999).
%
\bibitem{tsutsui}
K.~Tsutsui, T.~Tohyama, and S.~Maekawa 
Phy.~Rev.~Lett. {\bf 83}, 3705 (2000), 
and  
K.~Tsutsui, T.~Tohyama, and S.~Maekawa 
Phys.~Rev.~B {\bf 61}, 7180 (2000). 
%
\bibitem{hasan}
M.~Z.~Hasan, E.~D.~Isaacs, Z.~-X.~Shen, L.~L.~Miller, K.~Tsutsui, T.~Tohyama, and S.~Maekawa, 
Science  {\bf 288}, 1811 (1999).
%
\bibitem{inami}
T.~Inami, T.~Fukuda, J.~Mizuki, Y.~Murakami, K.~Hirota, and Y.~Endoh, 
Nucl. Instr. and Meth. (to be published).
%
\bibitem{ishihara1}
S.~Ishihara, and S.~Maekawa, 
Phy. Rev. B  {\bf 62}, 2338 (2000).
%
\bibitem{blume}
M.~Blume, 
J. Appl. Phys. {\bf 57}, 3615 (1985).
%
\bibitem{ishihara2}
S.~Ishihara, and S.~Maekawa,
Phys. Rev. B {\bf 62}, R9252 (2000).
%
\bibitem{forster}
D.~Forster, 
in {\it Hydrodynamic Fluctuations, Broken Symmetry and 
Correlation Functions} 
(Addison-Wesley, Massachusetts, 1990), and 
P.~Fulde, 
in {\it Electron Correlations in Molecules and Solids} 
(Spinger-Verlag, Berlin, 1993). 
%
\bibitem{slater}
J.~C.~Slater, and G.~F.~Koster, 
Phys.~Rev. {\bf 94}, 1498 (1954). 
%
\bibitem{saitoh}
T.~Saitoh, A.~E.~Bocquet, T.~Mizokawa, H.~Namatame, A.~Fujimori, M.~Abbate, Y.~Takeda, 
and M.~Takano, 
Phys. Rev. B {\bf  51}, 13942 (1995).
%
\bibitem{arima}
T.~Arima and Y.~Tokura, 
J.~Phys.~Soc.~Jpn.~{\bf 64}, 2488 (1995). 
%
\bibitem{ishihara0}
S.~Ishihara, J.~Inoue, and S.~Maekawa, 
Physica C {\bf 263}, 130 (1996), and Phys.~Rev.~B {\bf 55}, 8280 (1997).
%
\bibitem{hirota}
K.~Hirota, N.~Kaneko, A.~Nishizawa, Y.~Endoh, M.~C.~Martin, and G.~Shirane, 
Physica B {\bf 237-238}, 36 (1997). 
%
\end{references}
\end{document}